\documentclass[12pt]{article}

\usepackage{graphicx}
\usepackage{amsfonts}
\usepackage{amssymb}
\usepackage{amsmath}
\usepackage[colorlinks,urlcolor=blue]{hyperref}

\begin{document}

\title{\bf Quaternions in Hamiltonian dynamics of a rigid body -- Part III.\\
Asymmetric Top in the Orbitron}
\providecommand{\keywords}[1]{\textbf{\textit{Keywords: }} #1}

\author{
  Stanislav S. Zub\\
  Faculty of Cybernetics,\\
  Taras Shevchenko National University of Kyiv,\\
  Glushkov boul., 2, corps 6.,\\
  Kyiv, Ukraine 03680\\
  \texttt{stah@univ.kiev.ua}\\
  \\
  Sergiy I. Zub\\
  Institute of Metrology,\\
  Mironositskaya st., 42,\\
  Kharkiv, Ukraine 61002\\
  \texttt{sergii.zub@gmail.com}}

\maketitle

\newpage
\begin{abstract}
The Poisson structure in the quaternion variables was proposed
for asymmetric top in the external axially symmetric magnetic field.
For that model of interaction the motion equation were got.
The model was simulated in the neighbourhood of a relative equilibrium
of the symmetric top -- Orbitron by Monte Carlo method.
For small deviations with respect to the symmetry
we observe the persistence of the stability.

.

\end{abstract}

\newpage
\tableofcontents

%Here goes some text in koi8-r input encoding.
% To switch the current language (and font encoding) to English,
% use \English macro (or \Eng)
% To switch the current language (and font encoding) to Russian,
% use \Russian macro (or \Rus)
% Switching languages will, in particular, set the correct hyphenation.
% You may use several different input encodings in one document! :-)
%
%\inputencoding{cp1251}
% Вводим обозначения для матриц Паули (без шляпки, но жирно)
\newcommand{\s}[1]{\ensuremath{\boldsymbol{\sigma}_{#1}}}
% Вводим команду для жирного написания символов
\newcommand{\bsym}[1]{\ensuremath{\boldsymbol{#1}}}
% Вводим команду для внешнего умножения
\newcommand{\w}{\ensuremath{\boldsymbol{\wedge}}}
% Вводим команду для левой свертки
\newcommand{\lc}{\ensuremath{\boldsymbol{\rfloor}}}
% Вводим команду для правой свертки
\newcommand{\rc}{\ensuremath{\boldsymbol{\lfloor}}}

\thispagestyle{empty}

\newpage

\bigskip
\section{ Introduction }

\bigskip
In papers [1,2] were present the basic theoretical results on the investigation of use quaternions in Hamiltonian dynamics of a rigid body and in particular in the framework of the Poisson structure for quaternion variables we have the brackets
\[
\begin{cases}
   \{x_i,x_k\} = 0, \quad \{x_i,q_\mu\} = 0, \quad \{x_i,{\mathrm \Pi}_j\} = 0;\\
   \{p_i,q_\mu\} = 0, \quad \{p_i,{\mathrm \Pi}_j\} = 0;\\
   \{x_i,p_k\} = \delta_{ik}, \quad i, k = 0,1,2;\\
   \{q_\mu,q_\nu\} = 0, \quad \mu,\nu = 0,1,2,3; \\
   \{{\mathrm \Pi}_i, q_0\}  =  q_i;\\
   \{{\mathrm \Pi}_i, q_j\} = -q_0\delta_{ij} - \varepsilon_{ijl}q_l;\\
   \{{\mathrm \Pi}_i, {\mathrm \Pi}_j\}  = -2 \varepsilon_{ijl}{\mathrm \Pi}_l
\end{cases}
\leqno(1)\]
with quite general form of Hamiltonian:
\[ H\left((\bsym{x},\bsym{p}), (q, {\bsym{\mathrm \Pi}})\right)
   = \frac1{2 m}\bsym{p}^2
   + T_{spin}\left((\bsym{x},\bsym{p}), (q, {\bsym{\mathrm \Pi}})\right)
   + V(\bsym{x}, q)
\leqno(2)\]
where
\[ T_{spin}\left((\bsym{x},\bsym{p}), (q, {\bsym{\mathrm \Pi}})\right)
   = \frac18 {\bsym{\mathrm \Pi}} \mathbb{I}^{-1}{\bsym{\mathrm \Pi}}
   = \frac18 \left(\frac{{\mathrm \Pi}_1^2}{I_1} + \frac{{\mathrm \Pi}_2^2}{I_2} + \frac{{\mathrm \Pi}_3^2}{I_3}\right)
\leqno(3)\]
($\mathbb{I}$ -- the diagonal matrix of the tensor of inertia in the coordinates system associated with a body,
where corresponding diagonal elements are $I_1,I_2,I_3$; ${\mathrm \Pi} = 2{\mathrm M}$,
where ${\rm M}_i$ --  components of the intrinsic angular momentum in body frame)
the motion equations of a rigid body were obtained [1]:
\[
\begin{cases}
   \dot{\bsym{x}} = \frac1{m}\bsym{p};\\

   \dot{\bsym{p}} = -\partial_{\bsym{x}}V;\\

   \dot{q} = \frac12 q\bsym{\Omega}\longrightarrow q^{-1}\dot{q} = \frac12\bsym{\Omega};\\

  \dot{{\bsym{\mathrm \Pi}}}
   = -\Im\left( \bsym{\Omega}\bsym{\mathrm \Pi}
  + q^{-1}\nabla^{(q)}V\right)
\end{cases}
\leqno(4)\]
where $\Im(\ )$ --- vectorial part of the quaternion;
$x_i$ -- the coordinates of the center of mass a rigid body,
$q=(q_0,\bsym{q})$ -- the unit quaternion that describes the rotation from the inertial reference frame
to the frame of reference connected with a body,
$p_i$ -- the components of the linear momentum,
$\Pi_i$ -- the components of the angular momentum in the frame of reference associated with the body.

Here the translational degrees of freedom are given in the inertial reference frame,
and rotational in the reference frame associated with the body,
i.e.  we have ``mixed representation''.

It is interesting to apply the equations (4) to the solve of substantial physical problem.

For example, lets consider the problem of stability of the quasi orbital motions of a small magnetized
rigid body in an axially symmetric magnetic field [3,4,5].
In the cited papers was assumed that a small body with magnetic moment is a symmetric top
(i.e. $I_1=I_2=I_{\bot}$) in terms of mechanics.
And we applied the group-theoretic methods of Hamiltonian mechanics for investigations of them [4,6,7].
For so-called relative equilibria the authors succeeded in obtain an analytical proof of the motion stability.

Notice that a symmetrical rigid body is an idealized case,
so  it is important to know whether the stability is retained for the small deviations from symmetry,
i.e. for $I_1\neq I_2$.
But in that case  the group-theoretical approach that previously used becomes an unapplicable.
The direct numerical modeling of equations (4) is an alternative to an analytical study.

From a physical point of view, reasonable to expect that for small deviations
of the body parameters from a symmetric top and small deviations of initial conditions
from initial conditions of relative equilibrium a perturbation of the trajectory
will be also small and so stability that was detected analytically will remain unchanged.

{\it Remark.}
From a philosophical point of view, any experiment,
i.e. natural or numerical simulation can not prove the system stability.
They only can gives an arguments in favor of an existance of the stability.

From [8,9] known that description of a rigid body motion in quaternion variables
are substantially increases the efficiency of numerical simulation in comparison
with a description in matrices.

Among other things in contrast to the Euler angles the description in quaternion variables is invariant
and not required the computation of trigonometric function for the variables.
Beside that the description has only one redundant parameter in contrast of matrix description (see [6,10]).
Indeed the only one equation associates the four parameters of a quaternion vs matrix description
when orthogonality supported by using six constrains between its elements.
\[
q_0^2+q_1^2+q_2^2+q_3^2=1
\leqno(5)\]

However, even condition (5) leads to the problems at the numerical simulation of the equations (4).
This problem is well known and usually solved by creation of special geometric integrators [11].
In particular the corresponding integrator based on use of orthogonal matrices for description of
the rigid body dynamics was proposed in paper [10].

Mentioned above the advantage of using quaternion variables spreads
on their using in the integrator that is proposed here.

\bigskip
\section{ Mathematical model }

\bigskip
Our model consists from the following elements.

1. Magnetic field of the Orbitron.

Two opposite magnetic poles $\pm \kappa$ will be placed on $z$-axis at $\mp h$ points.
Thus the magnetic field of the Orbitron $\bsym{B}$ has the form:
\[ \bsym{B}(\bsym{r})
 = \sum_{\varepsilon=\pm 1} \bsym{B}_{\varepsilon}(\bsym{r}), \qquad
   \bsym{B}_{\varepsilon}
 = -\frac{\mu_0}{4\pi}\varepsilon\kappa
   \frac{\bsym{r}-\varepsilon h \bsym{e}_z}{|\bsym{r}-\varepsilon h \bsym{e}_z|^3}
\leqno(6)\]

By its construction the field $\bsym{B}$ has axial symmetry with respect to $z$-axis.

2. A small rigid body that interacts with a magnetic field is a magnetic dipole.
Simultaneously its mechanical properties correspond to an asymmetrical body with mass $m$
and the principal moments of the inertia $I_1,I_2,I_3$.
Let $|\vec{\mu}|{\rm N}$ --- magnetic moment (where {\rm N} is 3-component {\it arithmetical vector})
of the rigid body is a fixed vector in the system associated with the body.

{\it Remark}. In general its direction is not coincide with any of the directions of the principal axes of inertia
in contrast to the symmetric top, where we assumed that the magnetic moment parallel to the vector $\vec{E}_3$.

3. We obtain the Hamiltonian equations of motion for magnetic dipole-top in the external magnetic field from (4) with Hamiltonian (3).

Let's write the potential energy of the system and
its derivatives with respect to the spatial and quaternion variables explicitly.

\bigskip
Potential energy of a dipole in a magnetic field is
\[ V(\vec{x},q) = -\langle\vec{\mu},\vec{B}\rangle
   = -|\vec{\mu}|\langle\vec{\nu},\vec{B}\rangle
\leqno(7)\]

Let $\vec{e}_i$ are unit vectors of inertial system and
$\vec{E}_k$ are directing unit vectors of the principal axes of inertia
and $Q_{ik}$ matrix transformation the variables in the body frame
to the variables in the inertial system, i.e.
\[ \vec{E}_k = Q_{ik}\vec{e}_i
\]

Matrix $Q_{ik}$ in quaternion variables have the form
\[  Q_{ik} = (2q_0^2 - 1)\delta_{ik}
    + 2 q_i q_k - 2 q_0 q_j\varepsilon_{jik}
\]

Relation between components ${\rm N}_k$ of the directing unit vector of the magnetic moment
in the body frame and in the inertial system is $\nu_i$ has the form
\[ \nu_i = Q_{ik}{\rm N}_k
\leqno(8)\]

Then
\[  \langle\vec{\nu},\vec{B}\rangle
  = \nu_i B^i = B^i Q_{ik}{\rm N}_k
\leqno(9)\]
\[
  = (2q_0^2 - 1) ({\rm N}_i B^i)
    + 2 ({\rm N}^k q_k )(q_i B^i)
    + 2 q_0 q_j\varepsilon_{jki}{\rm N}^k B^i
\]

We can give an invariant form for the expression (9).

Let's suppose that ${\rm N}^k$ and $B^i$ are the components of the pure quaternions
(but in geometric sense they are the components of the vectors in the {\it different} frames of reference).

By using quaternions the expression (8) can be written in form
\[  \bsym{\nu} = q\bsym{{\rm N}}q^{-1}
\leqno(10)\]

Accordingly
\[  \langle\vec{\nu},\vec{B}\rangle
    = \langle\bsym{B}, q\bsym{{\rm N}}q^{-1}\rangle
    = \langle\bsym{B} q, q\bsym{{\rm N}}\rangle
\leqno(11)\]

Then the total differential with respect to $q$ has the form
\[ d^{(q)}\langle\vec{\nu},\vec{B}\rangle
   = d^{(q)}\langle\bsym{B} q, q\bsym{{\rm N}}\rangle
   = \langle\bsym{B} d q, q\bsym{{\rm N}}\rangle
   + \langle\bsym{B} q, d q\bsym{{\rm N}}\rangle
\leqno(12)\]
i.e.
\[ d^{(q)}\langle\vec{\nu},\vec{B}\rangle
   = \langle\bsym{B} e_\mu d q^\mu , q\bsym{{\rm N}}\rangle
   + \langle\bsym{B} q, e_\mu d q^\mu\bsym{{\rm N}}\rangle
\leqno(12a)\]

Hence
\[ \frac{\partial}{\partial q^\mu}\langle\vec{\nu},\vec{B}\rangle
   = \langle\bsym{B} e_\mu, q\bsym{{\rm N}}\rangle
   + \langle\bsym{B} q, e_\mu\bsym{{\rm N}}\rangle
\leqno(12b)\]

{\it Remark}. For the scalar product of quaternions the following representations are true
\[ \langle a, b\rangle = \frac12(a b^\dag + b a^\dag)
  = \langle a^\dag, b^\dag\rangle = \frac12(a^\dag b + b^\dag a)
\leqno(13)\]
The important properties of the scalar product are arise from here
\[
\langle q a, b\rangle = \langle a, q^\dag b \rangle;
\quad
\langle a, b q\rangle = \langle a q^\dag, b \rangle,
\leqno(14)\]
where $q,a,b$ -- arbitrary quaternions.

Then we have ($\bsym{B}^\dag = -\bsym{B}$ and $\bsym{\rm N}^\dag = -\bsym{\rm N}$)
\[ \nabla^{(q)}\langle\vec{\nu},\vec{B}\rangle =
   e_{\mu}\langle\bsym{B} e_{\mu}, q\bsym{{\rm N}}\rangle
   +  e_{\mu}\langle\bsym{B} q, e_{\mu}\bsym{{\rm N}}\rangle =
\leqno(15)\]%(13)
\[ -  e_{\mu}\langle e_{\mu}, \bsym{B}q\bsym{{\rm N}}\rangle
   -  e_{\mu}\langle\bsym{B} q\bsym{{\rm N}, e_{\mu}}\rangle =
\]
\[ -  2e_{\mu}\langle e_{\mu}, \bsym{B}q\bsym{{\rm N}}\rangle =
   -  2\bsym{B}q\bsym{{\rm N}}
\]
and
\[ q^{-1}\nabla^{(q)}\langle\vec{\nu},\vec{B}\rangle =
   - 2(q^{-1}\bsym{B}q)\bsym{{\rm N}}
\leqno(16)\]%(14)
analogically from (11) by differentiating with respect to $\vec{x}$ we obtain
\[ \dot{p}_i = -\partial_{\bsym{x}}V = |\vec{\mu}|\langle q^{-1}\bsym{{\rm N}}q, \partial_{i}\bsym{B}\rangle
\leqno(17)\]%(15)

Substituting (16) and (17) into (4) we obtain equations of motion for the dipole-top
\[
\begin{cases}
\dot{\bsym{x}} = \frac1{m}\bsym{p};\\

\dot{p}_i = |\vec{\mu}|\langle q^{-1}\bsym{{\rm N}}q, \partial_{i}\bsym{B}\rangle;\\

\dot{q} = q \frac{\bsym{\Omega}}{2}, \quad \Omega_i = \frac12 (I^{-1})_{ik} {\mathrm M}_k;\\

\dot{{\bsym{\mathrm \Pi}}}
= -\Im\left( \bsym{\Omega}\bsym{\mathrm \Pi}
  + 2|\vec{\mu}|(q^{-1}\bsym{B}q)\bsym{{\rm N}}\right)\\
\end{cases}
\leqno(4a)\]

\bigskip
\section{ Geometric integrator }

\bigskip
In the article [10] was declared the importance of applying the geometric integrators
in study of the systems with energy conservation on long time intervals [11].
It is equally important that in our case not only the Poisson structure conserved
but also the Casimir function (5).
Violation of the last condition means the exit of the scope of the theory of rigid body.

In contrast to [10] where the matrix approach to description of a rigid body was implemented
here we use quaternion approach.
From one hand that is simplifies the control of orthogonality of a movable basis
of an asymmetric rigid body but other hand requires finding of an analytical solution
of the third equation of the system (4a).

Similarly to [10] for solution of (4a) on one integration step are realized the consecutive integration
of equations with potential and kinetic energies.
Moreover the part of system (4a) with the kinetic energy integrates fully analytically.
But for the integration of part related to potential energy is applied numerical method of integration
of the 2-nd order as in [11].

The analytical solution of the equation
\[
\dot{q} = q \frac{\bsym{\Omega}}{2}
\]
where $\bsym{\Omega} = const$ --- pure quaternion is given by
\[ q(t) = q(t_0) \exp \left(\frac{\bsym{\Omega}(t-t_0)}{2}\right)
\leqno(18)\]%(15)

{\it Remark}. Function $exp$ is completely determined by the following properties
\[
\begin{cases}
   \exp(\frac{\bsym{\Omega}}{2} t_1) \exp(\frac{\bsym{\Omega}}{2} t_2) = \exp(\frac{\bsym{\Omega}}{2} (t_1 + t_2));\\
   \exp{(0)} = 1.
\end{cases}
\leqno(19)\]%(16)

From (19) is easy to verify
\[\exp \left(\frac{\bsym{\Omega}(t-t_0)}{2}\right)
  = \cos\left(\frac{|\bsym{\Omega}|(t-t_0)}{2}\right) +
\sin\left(\frac{|\bsym{\Omega}|(t-t_0)}{2}\right)\frac{\bsym{\Omega}}{|\bsym{\Omega}|}
\leqno(20)\]%(17)

Accordingly the solution of equation
\[
\dot{\bsym{x}} = \frac1{m}\bsym{p}
\]%(18)
has the form
\[ \bsym{x}(t) = \bsym{x}(t_0) + \frac{\bsym{p}(t_0)}{m}(t-t_0)
\leqno(21)\]%(19)

Thus (18) and (21) give the expressions for our geometric integrator in quaternion representation.

\bigskip
\section{ Estimation of physical parameters }

\bigskip
As already mentioned above, as a starting point,
i.e. the set of physical parameters and initial conditions
we take the relative equilibrium in the Orbitron system for a symmetric top [3].
For this system the realistic physical parameters
based on the properties of modern magnetic materials were choosen.

These parameters presented bellow.

For the magnets produced from $NdFeB$
density --- $\rho = 7,4 \cdot 10^3(kg/m^3)$ and
residual magnetic induction --- $B_r = 0,25 (T)$.
Then easily to get the «charge» of magnetic pole $\kappa = 17,6 (A \cdot m)$.
The distance between poles $L = 2h = 0,1 (m)$.

Let's chose the movable magnet in the form of a cylinder (or rather disk)
with a diameter $d = 0,014 (m)$ and height $l = 0,006 (m)$.
So, the magnetic moment of the disk $\mu~=~0,18 (A \cdot m^2)$.

Eventually for the orbit of radius $r_0 = 1,5 h = 0,075 (m)$
we obtain the angular velocity of the orbital motion $\omega = 1,54 (rad/sec)$,
at that the minimum angular velocity of the disk rotation in this case is $\Omega = 72,8 (rad/sec)$.
Such values of the angular velocities look quite reasonable.

\bigskip
{\it Remark}. The initial conditions for the relative equilibrium of the Orbitron
($I_1=I_2=I_{\bot}$, ${\rm N} = (0,0,1)$ --- in body frame):

1.1. Translational variables --- the same values as for Orbitron.

1.2. Rotational variables $q = e_0 = (1,0,0,0)$.

1.3. Initial value ${\mathrm \Pi} = 2{\mathrm M}$, where ${\rm M}$ is initial value for Orbitron.

1.4. At start point the base vectors of the inertial system and body frame are coincide.

\bigskip
Variations of the initial conditions are analogous to the Orbitron.
In addition to variations of the initial conditions the subject to variations
are the parameters of the body, namely, the moments of inertia $\mathbb{I}$
and directing unit vector of the magnetic moment ${\rm N}$.

\bigskip
{\it Remark}: The quaternions $q, {\rm N}$ must to be normalized after operations of variations.

\bigskip
We conducted a simulation of 10000 trajectories that differ from each other
with the random variations of parameters and
initial conditions in the range of relative deviations from the starting point $\sim 1\%$.
Each of the trajectories consistes of 100 quasi-orbits.
Stability loss has not been detected for these quasi-orbits.

Despite the fact that first stable orbital motions have been found based
on group-theoretic methods in mechanics for systems with symmetry [3,5,12],
the results of this work show that with respect to the problem of dynamic stability
the requirements of the symmetry are not critical.
Indeed, at least for small deviations with respect to the symmetry
we observe the persistence of the stability.

\end{document}